\documentclass[final]{statsoc}

\usepackage{times}
\usepackage{bm}
\usepackage{natbib}
\usepackage{amsmath,amssymb}
\usepackage{amsfonts}
\usepackage{tikz}
\usepackage{graphicx}
\usetikzlibrary{matrix,arrows}
\usepackage{subfigure}

\usepackage[plain,noend]{algorithm2e}

\def\P{{\cal P}}

\def\pr{{\text{pr}}}

\def\PE{\mathop{\text{Ewens--Pitman}}\nolimits}

\def\Nb{\mathop{\mathbb{N}_{}}\nolimits}

\numberwithin{equation}{section}

\received{\today}

\title{Community detection for interaction networks}

\author{Harry Crane}
\address{ Department
of Statistics \& Biostatistics, 
 Rutgers University, 
110 Frelinghuysen Road, 
Piscataway, NJ 08854, USA
 \email{hcrane@stat.rutgers.edu}}
\author{Walter Dempsey}
\address{ Department
of Statistics, 
University of Michigan,
1085 S.\ University Avenue,
Ann Arbor, MI 48109, USA
 \email{wdem@umich.edu}}

\begin{document}

\maketitle

\begin{abstract}
In many applications, it is common practice to obtain a network from interaction counts by thresholding each pairwise count at a prescribed value.
Our analysis calls attention to the dependence of certain methods, notably Newman--Girvan modularity, on the choice of threshold.
Essentially, the threshold either separates the network into clusters automatically, making the algorithm's job trivial, or erases all structure in the data, rendering clustering impossible.
By fitting the original interaction counts as given, we show that minor modifications to classical statistical methods outperform the prevailing approaches for community detection from interaction datasets.
We also introduce a new hidden Markov model for inferring community structures that vary over time.
We demonstrate each of these features on three real datasets: the karate club dataset, voting data from the U.S.\ Senate (2001--2003), and temporal voting data for the U.S.\ Supreme Court (1990--2004).
\end{abstract}


\section{Introduction}\label{section:introduction}

Networks represent dependencies and interactions among individuals, genes, and particles in diverse social, biological, and physical science applications.
The sheer complexity of network datasets presents conceptual and computational issues that often limit the availability of practical measures for extracting meaningful information.
The wealth of literature on community detection attempts to tame this complexity by dividing the network into clusters (or communities) of vertices, with the hope that this community structure provides a sparse or low resolution representation of the network.
Heuristically, vertices within the same cluster can be regarded as interchangeable, and network structure for $n$ vertices, and therefore $O(n^2)$ interactions, is effectively parameterized by a much smaller number of communities.

Empirical evidence suggests that this approach works well in practice, and recent mathematical results by \citet{ZhaoLevinaZhuconsistency2011} make this heuristic rigorous in the case of the stochastic blockmodel (SBM) \citep{HollandLaskeyLeinhardt1983}.
Various alternative approaches and refinements, e.g., degree-corrected stochastic blockmodels \citep{KarrerNewman2011}, mixed membership models \citep{AiroldiBleiFienbergXing2008}, modularity-based algorithms \citep{BickelChen2009PNAS,NewmanGirvan2004,ZhaoLevinaZhu2011}, and spectral clustering algorithms \citep{spectral}, have been proposed for treating heterogeneous networks.

Despite these efforts, the development of principled and practical statistical methods has been slow relative to the explosive growth in the field of {\em network science} over the past twenty years.
A major obstacle is added uncertainty about how the observed network data relates to the real world phenomenon of interest.
Various authors have demonstrated the drastic effect of sampling on network data \citep{LeeKimJeong2006,WillingerAldersonDoyle2009}, calling into question whether the  ``scale-free'' behavior observed by \citet{BA1999} and several others is a real phenomenon or merely an artifact of sampling.
In an effort to understand this added uncertainty in network modeling, \citet{CraneDempsey2015} have demonstrated that in general a statistical network model should reflect both the network generating process and the sampling mechanism used to produce the observed network data.

In most cases, the sampling mechanism is not understood well enough to nicely incorporate into a statistical model, and so we shall not address that important problem of network analysis here.
Instead, we demonstrate the issues of network sampling in a simple, but commonly encountered, setting.
Many ``network datasets'' are derived from much richer datasets that involve interactions among a group of individuals.
In this case, the interaction data contain much more information than a boolean indicator of a relationship between each pair of individuals; they may also contain a count for the number of interactions over a given period, or a frequency of interaction, or some other covariate information.
The karate club dataset \citep{Zachary1977} is a canonical example.
Here the full dataset is a two-way table of interaction counts between all individuals in a karate club.
Most analyses of this dataset, however, work with the projected array of boolean indicators, with a 1 indicating that the corresponding pair of individuals had at least one interaction and a 0 indicating otherwise.
Innocuous as it appears, we demonstrate the sensitivity of prior methods to the choice of threshold in the karate club dataset (Section \ref{section:karate-application}) and two other datasets of voting behavior in the U.S.\ Senate and U.S.\ Supreme Court (Sections \ref{section:Senate-application} and \ref{section:supreme court-application}).

\citet{KarrerLevinaNewman2008} have previously studied robustness of certain algorithmic methods by measuring the {\em variation of information} with respect to random perturbations in network structure.
Their measure appears to work well in discerning those networks which possess a strong community structure, but it does not address the more preliminary issue of robustness to the network sampling mechanism.
This latter issue is rarely raised in network applications, but here we demonstrate the strong dependence (and therefore lack of robustness) of modularity-based approaches on the sampling procedure.
Though different than the variation of information criterion, variation in the sampling mechanism is a very real obstacle faced by methods fit to thresholded network data.
The sensitivity of the prevailing Newman--Girvan modularity to the choice of threshold (Figures \ref{figure:2} and \ref{figure:3}, Table \ref{table:percentile}) underscores a key point of \citet[p.\ 2]{KarrerLevinaNewman2008}, ``If a small change in the network---an edge added
here, another deleted there---can completely change the outcome
of our community finding calculations then, we argue,
the communities found should not be considered trustworthy.''

We further explore the extent to which the act of projecting interaction data to a network can be avoided altogether by simply modeling the interaction table as is, eliminating any concern over how the projection was chosen.
With this, we show that readily available statistical methods outperform the prevailing network methods in each of our three data examples.
As a particularly illuminating example, we show that a simple two parameter Poisson model for interaction counts exactly recovers the known community structure in the karate club network, while the degree-corrected stochastic blockmodel \citep{KarrerNewman2011} requires thirty-six parameters and still incorrectly specifies one individual; see Section \ref{section:karate-application}.
In modeling the data as it comes, we avail ourselves to many techniques from classical statistics, allowing us to easily interpret model output and arming our approach with considerable flexibility to handle a range of clustering problems.
We demonstrate this latter point with a real data example for cluster detection in temporally varying networks; see Section \ref{section:supreme court-application}.


\section{Motivating examples}\label{section:examples}

We frame our discussion around three real data examples from the social and political science literature: the karate club \citep{Zachary1977}, senate voting  \citep{Crane2014JASA}, and supreme court voting\footnote{accessed at http://scdb.wustl.edu/index.php} datasets.
The karate club dataset is the canonical example for community detection in networks.
The senate dataset was introduced by \citet{Crane2014JASA} in the context of clustering from categorical data sequences, and here we introduce it in the realm of network analysis.
The supreme court dataset consists of all U.S.\ Supreme Court (USSC or `the Court') decisions during a fifteen year span of the Rehnquist Court (1990--2004); we use it to illustrate the potential for certain partition-valued Markov chains in modeling temporal clustering as well as to highlight the issues faced by other approaches in the presence of time-varying data.
Each of these examples highlights a different feature of community detection, as we outline in Section \ref{section:summary}.


\subsection{Karate club dataset}\label{section:karate}

Zachary's \citep{Zachary1977} karate club dataset records the number of social interactions of thirty-four members in a karate club that experienced a split between its two leaders.
Because the resulting split of the members into two groups is well understood, the so-called karate club dataset is the canonical example for community detection in network data.
\citet[Figs.\ 2 \& 3]{Zachary1977} records both the interaction counts and the unweighted network with edges representing those pairs of individuals with a positive interaction count.
The resulting dataset is a square array with 34 rows and columns corresponding to the thirty-four members of the karate club.

\subsection{Senate dataset}\label{section:senate}

\citet{Crane2014JASA,Crane2014Biometrika} analyzed voting alignments for every bill in the 107th U.S.\ Senate (2001--2003) by treating the outcome of each bill as an independent, identically distributed (i.i.d.) draw from a partition model for categorical responses.
In this context, the clustering should reflect the political allegiances of the 100 senators.

By considering each vote outcome separately, those prior analyses manage to simultaneously incorporate two-way, three-way, and higher-order interactions among senators.
Here we summarize the voting alignments more simply in terms of two-way interaction counts, i.e., $A_{ij}=(N_{ij},V_{ij})$ records the number of votes $V_{ij}$ on which senators $i$ and $j$ agreed out of $N_{ij}$ bills on which they both voted.
Our approach simplifies the dataset considerably while yielding the same insights; however, the same is not true for other leading methods when fit to the projected network.

Rather dramatically, the senate dataset demonstrates the fickle nature of projecting interaction counts to an unweighted network.
Over the course of the 107th U.S. Senate term, every pair of senators voted in agreement at least once, in fact hundreds of times.
In this context, it is more natural to threshold based on the proportion $V_{ij}/N_{ij}$ of time senators $i$ and $j$ agreed; however, the many possible choices of this cutoff value leave considerable influence in the hands of the data analyst.
Our analysis in Figure \ref{figure:2} and Table \ref{table:percentile} point out the lack of robustness of a leading method, Newman--Girvan modularity \citep{BickelChen2009PNAS,NewmanGirvan2004}, to this choice of cutoff.


\subsection{Supreme Court dataset}\label{section:supreme court}

The Supreme Court interaction dataset has the same form as the Senate dataset, with the added feature of a temporal collection of interaction arrays over the years 1990--2004.
On each of about 80 cases per year, the nine Supreme Court justices rule for one of the two sides.
Justices declare no official political or ideological allegiances, but their philosophy and personal views are well documented and we expect the clustering to reflect this separation.
The interaction array for a given term $t$ records the number of times the justices voted in agreement, i.e., $A^t=(A^t_{ij})_{1\leq i,j\leq 9}$ with $A^t_{ij}=(N^t_{ij},V^t_{ij})$ keeping track of how many times justices $i$ and $j$ agreed ($V^t_{ij}$) and how many cases they both ruled ($N^t_{ij}$) during term $t$.
The dataset records these data for the judicial terms $t=1990,\ldots,2004$.
The collection $(A^t)_{t=1990,\ldots,2004}$ records these interaction arrays over time, and we are interested in detecting changes to the Court's ideological alignment during this period.

Models that allow for temporally varying communities in networks are important for detecting regime change in political and social science datasets.
This represents an underdeveloped area with only a few attempts at establishing a viable framework \citep{HuhFienberg2008,Snijders2006}.
Using recently developed theory from the literature on  partition-valued Markov chains \citep{Crane2011a,Crane2014AOP}, we model community dynamics in the above supreme court dataset with a hidden Markov chain on partitions.
Although the Court's membership is not constant over the period we study, special properties of the chosen Markov model nullify these issues, producing sound inferences; see Section \ref{section:supreme court-application}.

\subsection{Summary of analysis}\label{section:summary}

Each of the above examples illustrates a different aspect of modeling interaction data.
The karate club analysis puts our methods on equal footing with prior approaches by showing that it performs as well (and in fact a bit better) than many prevailing techniques.
The senate dataset allows us to further explore the effect of projecting data on interaction counts to a network without edge weights.
The time period we study for the Supreme Court (1990--2004) has been examined previously in  legal studies \citep{Toobin2008} and also quantitative political science \citep{Sirovich2003PNAS}, but here we introduce it as an example of how to detect changes in network clustering over time.
For this, we bring over some recent developments in the theory of partition-valued Markov chains \citep{Crane2011a}.


Without the need for degree-correction or other sophisticated techniques, 
we show that straightforward modifications of classical statistical methods fit to 
the \emph{observed} data
outperform the community detection methods put forth by \cite{BickelChen2009PNAS} and \citet{KarrerNewman2011}.
To a large extent, many of our models are not at all new---our hidden Markov model for community detection in temporally varying networks is a novel contribution---but they do entail some subtle considerations of network data that have not been given much attention.
Perhaps most significant is our thorough testing of the often-overlooked effect of sampling on network analysis, which provides a cautionary tale about misinterpreting inferences from certain state-of-the-art methods.
At the very least, our analysis reiterates that Occam's razor---the simplest explanation is often best---applies just as well to network modeling.


\section{Interaction data}\label{section:data}

All of the above datasets arise by repeated {\em interactions} among a population of individuals.
The karate club dataset contains counts of the number of social interactions outside of the club during a specific period of time; in the senate, counts are the number of bills on which the senators voted in agreement during the 107th congressional term; and in the USSC, interactions entail judicial decisions on which two justices agreed, with an array of interaction counts for each of the fifteen judicial terms between 1990 and 2004.
Each array of interaction counts gives rise to a network by projecting, in a number of possible ways, to an array of $\{0,1\}$-valued indicators.

We acknowledge the extensive literature on modeling relational data in economics, social and biological sciences, e.g., \citep{Bergmann2003,Lazzarini2001}; however, many of these methods deal explicitly with normal data \citep{LiLoken2002} and other data forms \citep{Hoff2005,Hoff2008}.
Other methods, such as latent space models \citep{HoffRafteryHandcock2002}, seem amenable to network analysis, but we do not pursue these here.
If at all possible, we favor the simplest model that makes sense for the given application, reaping the benefits of clarity when interpreting the inferred clustering.

\subsection{General setup}
All of the datasets above have the form of an array generated by repeated interactions within a population.
We observe data for a finite sample $\mathcal{S}$ from a finite or countable population of individuals $\P$.
For notational convenience, we label the population with the positive integers $\mathbb{N}=\{1,2,\ldots\}$ and we identify $\mathcal{S}$ as the first $n$ of these $\mathcal{S}=[n]:=\{1,\ldots,n\}$.
The population clusters into non-overlapping classes according to a partition $B=\{B_1,B_2,\ldots\}$, where $B_1,B_2,\ldots$ are non-empty, disjoint, and satisfy $\bigcup_{i\geq1}B_i=\P$.
The response for a sample $\mathcal{S}=[n]:=\{1,\ldots,n\}$ takes the form of an {\em interaction array} $A=(A_{ij})_{1\leq i,j\leq n}$, where $A$ takes values in some space $\mathcal{A}$ so that $A_{ij}$ reflects the strength of interaction or relationship among individuals~$i$ and $j$ in the sample.
In the examples we consider, $A_{ij}$ counts the number of interactions of a single type (Section \ref{section:karate}) or contains information about interactions of different types, such as {\em agree} and {\em disagree} (Sections \ref{section:senate} and \ref{section:supreme court}).

In networks applications, it is common to reduce the information in $A$ to an {\em adjacency array} $A^*_{c}=(A^*_{ij,c})_{1\leq i,j\leq n}$, where $A^*_{ij,c}=1$ if and only if $t(A_{ij})> c$ for some chosen {\em cutoff} $c\geq0$ and a {\em thresholding function} $t:\mathcal{A}\to[0,\infty)$ that combines the information in $A_{ij}$.
For example, in projecting the karate club dataset to an adjacency array, \citet[Fig.\ 2]{Zachary1977} implicitly uses the threshold $c=0$ and the identity function $t(A_{ij})=A_{ij}$.
(The senate and supreme court datasets have $A_{ij}=(N_{ij},V_{ij})$, for which the proportion $t(A_{ij})=V_{ij}/N_{ij}$ is a natural thresholding quantity.)
From now on, we use the term {\em network data} to generically refer to interaction data.
We refer to the adjacency array  $A^*_c$ as the {\em projected network}, which has vertex set $\mathcal{S}$ and edge set $E\subseteq \mathcal{S}\times\mathcal{S}$ satisfying $(i,j)\in E$ if and only if $A^{*}_{ij,c}=1$.

\subsection{Interpreting the data}\label{section:interpretation}

Simple data generating models typically lead to interpretable inferences and clear insights for relational data, which commonly arise in applications with pairwise measurements on the observed sample.
As mentioned above, the clarity of this interpretation can be obscured by the act of projecting $A\mapsto A^*_c$ to an unweighted network.
In \citet{CraneDempsey2015}, we advised against thresholding on logical grounds, but our analysis here demonstrates its drawbacks empirically.

We emphasize these consequences because thresholding is a common approach to obtain a projected network from interaction data in social sciences, where the network $A^*$ obtained by putting $A^*_{ij,c}=\mathbf{1}_{\{A_{ij}>c\}}$ has the interpretation of a social network where $i$ and $j$ are friends if they have interacted more than $c$ times within some prespecified period of time.
In principle, the distribution of this projection can be determined, but there are some subtleties introduced by the fact that  the cutoff value is often chosen after looking at the data.
We see the effect of this throughout Section \ref{section:applications}.

As we will see, in the karate club dataset, the standard projection with cutoff $c=0$ leads to an inferred clustering with one wrongly classified individual under the Newman--Girvan (NG) modularity \citep{NewmanGirvan2004}, while the projection with cutoff $c=1$ leads to the correct clustering under NG modularity.
It may seem harmless enough in this simple setting of the karate club, but adverse effects of network sampling on inferences are well documented \citep{CraneDempsey2015,LeeKimJeong2006,WillingerAldersonDoyle2009} and one cannot be sure that inferences from sampled networks are truly reflective of the real world generating process.
Table \ref{table:percentile} demonstrates this lack of robustness of NG modularity for the senate voting data.
In this case, we see that the projection algorithm is more responsible for detecting the true clustering than the algorithm: either the cutoff value is well chosen, in which case the projection effectively separates the nodes into clusters without the algorithm's help, or the cutoff destroys the structure in the data, leaving the algorithm hopeless in discovering latent structure.  

\section{Modeling the interaction array}\label{section:generating process}

For the datasets we consider, we need not open ourselves up to the above edge sampling issue.
Instead, we opt to work with the full interaction array $A=(A_{ij})_{i,j=1,\ldots,n}$ in all our analyses.
There are many ways to model these data without projecting to $A^*_c$, and our choice naturally depends on the features of each application.
We encounter two situations in our examples: either $A$ represents interactions among individuals over some period of time, or $A$ represents a fixed number of interactions with each interaction having a type, e.g., {\em agree}, {\em disagree}, or {\em undetermined} in the senate and supreme court datasets.
We discuss each in turn.

\subsection{Interaction count data}\label{section:count data}
Consider the case where $A=(A_{ij})_{i,j=1,2,\ldots}$ consists of a single interaction count, $A_{ij}\in\Nb$ for every $i,j=1,2,\ldots$.
In the most generic setting, we let $\Lambda=(\lambda_{ij})_{i,j=1,\ldots,n}$ be a matrix of non-negative intensities $\lambda_{ij}\geq0$.
Given $\Lambda$, we assume $A$ results from a Poisson point process on $[n]\times[n]$ with intensity measure $\Lambda$, i.e., the counts $(A_{ij})_{1\leq i,j\leq n}$ are independent with each $A_{ij}\sim\text{Poisson}(\lambda_{ij})$.
For a given interaction array $(a_{ij})_{1\leq i,j\leq n}$, we have
\begin{equation}\label{eq:Poisson-SBM}
\pr(A=(a_{ij})_{1\leq i,j\leq n};(\lambda_{ij})_{1\leq i,j\leq n})=\prod_{1\leq i,j\leq n}{\lambda_{ij}^{a_{ij}}}e^{-\lambda_{ij}}/{a_{ij}!}.
\end{equation}
(Note that in the symmetric setting, $A_{ij}=A_{ji}$, we consider only the counts $(A_{\{i,j\}})_{1\leq i<j\leq n}$.)
We refer to this model as the {\em Poisson stochastic blockmodel} below.

For community detection, we assume the population clusters according to a partition $B=\{B_1,B_2,\ldots\}$, and we can allow the intensities $\Lambda$ to depend on $B$ in a similar fashion to the stochastic blockmodel.
In this way, we define $\lambda_{ij}=\Lambda(B(i),B(j))$, where $B(i)$ is the block of $B$ that contains $i$.
\citet{KarrerNewman2011} introduce both the stochastic blockmodel and its degree-corrected version in terms of Poisson counts, just as we have here.
However, they note that this is a matter of mathematical convenience, and it seems they have not taken full advantage of the added power of this approach as a model for the interaction counts directly.

For logical reasons, it may make sense to simplify the parameter space of the Poisson stochastic blockmodel further by specifying the intensity of all within-cluster interactions by a single parameter, and likewise for all between-cluster interactions.
The justification for this depends on the given application.
For example, if we {\em a priori} expect the interaction behaviors within different clusters to be similar, then it makes sense to choose the simplest available model by putting $\lambda_{ij}=\lambda_{\text{in}}$ if $i$ and $j$ are in the same block of $B$ and $\lambda_{ij}=\lambda_{\text{out}}$ otherwise.
This gives the resulting clustering a clear interpretation in terms of the specified model and avoids potential issues of overfitting.

Even in the absence of any intuition for the cluster behavior, the interest of elegance and parsimony suggest that it is best to cut down on additional parameters whenever possible, especially since the clustering $B$ is our main interest.
As a rule of thumb, \citet{McCullaghYang2008} suggest at most 5 parameters, and in our analysis we never need more than 2.
This is a stark contrast to the approach of the degree-corrected stochastic blockmodel, which in general has on the order of $n+k^2$ parameters for a sample of size $n$ and partition $B$ with $k$ clusters.

As we demonstrate in Section \ref{section:karate-application}, the Poisson stochastic blockmodel with two parameters $(\lambda_{\text{in}},\lambda_{\text{out}})$ fit to the interaction counts recovers the correct clustering in the karate club dataset without the need for degree-correction \citep{KarrerNewman2011} or other constraints \citep{BickelChen2009PNAS}.
The best known performance of these latter approaches incorrectly classifies one individual.


\subsection{Interactions with types}\label{section:types}

In the Senate and Supreme Court datasets, the interaction array $A=(A_{ij})_{1\leq i,j\leq n}$ includes more information than simply the number of interactions between senators or judges.
Here we interpret interactions in the context of bills voted (resp.\ cases ruled) on by the U.S.\ Senate (resp.\ U.S.\ Supreme Court), and we define an {\em interaction between senators (resp.\ judges) $i$ and $j$} as a bill (resp., case) on which the two senators (resp., judges) both voted (resp., ruled).
Each interaction, therefore, has a type {\em agree} and {\em disagree}, and we observe a pair $A_{ij}=(N_{ij},V_{ij})$ with $N_{ij}$ the number of interactions between $i$ and $j$ and $V_{ij}$ is the number of times they agreed.

It is natural to assume that ``non-interactions,'' i.e., bills or cases for which at least one of $i$ and $j$ was absent, occur completely at random and independently of the observed interactions.
Prior analyses of the Senate and the Supreme Court make these assumptions without any apparent ill effects; we expect the same here as such instances are rare relative to the overall number of interactions.
Given $N=(N_{ij})_{1\leq i,j\leq n}$, therefore, we model $V=(V_{ij})_{1\leq i,j\leq n}$ as independent Binomial random variables with success probabilities $(p_{ij})_{1\leq i,j\leq n}$, where each $V_{ij}\sim\text{Binomial}(N_{ij},p_{ij})$.
In general, the probability of a given observation $A=(a_{ij})_{1\leq i,j\leq n}$ based on $N=(n_{ij})_{1\leq i,j\leq n}$ and $(p_{ij})_{1\leq i,j\leq n}$ is
\begin{equation}\label{eq:binomial-SBM}
\pr(A=(a_{ij})_{1\leq i,j\leq n};(n_{ij})_{1\leq i,j\leq n},\ (p_{ij})_{1\leq i,j\leq n})=\prod_{1\leq i,j\leq n}\binom{n_{ij}}{a_{ij}}p_{ij}^{a_{ij}}(1-p_{ij})^{n_{ij}-a_{ij}}. 
\end{equation}
We incorporate clustering into the model just as for the above Poisson stochastic blockmodel by regarding $P:B\times B\to[0,1]$ as function on pairs of blocks and putting $p_{ij}=P(B(i),B(j))$.
We call this the {\em Binomial stochastic blockmodel}.

In the senate dataset below, $A$ is symmetric and we fit the simplified Binomial stochastic blockmodel with two parameters $p_{\text{in}},p_{\text{out}}\in[0,1]$ and
\[p_{ij}=P(B(i),B(j))=\left\{\begin{array}{cc}
p_{\text{in}},& i\text{ and }j\text{ in the same block of }B,\\
p_{\text{out}},& \text{otherwise.}
\end{array}\right.\]
In this case, the distribution in \eqref{eq:binomial-SBM} simplifies to
\begin{eqnarray}\label{eq:binomial-2par}
\lefteqn{\pr(A=(a_{ij})_{1\leq i,j\leq n};(n_{ij})_{1\leq i,j\leq n},B, p_{\text{in}},p_{\text{out}})=}\\
&=&\prod_{1\leq i<j\leq n}\binom{n_{ij}}{a_{ij}}p_{\text{in}}^{a_{ij}B(i,j)}p_{\text{out}}^{a_{ij}(1-B(i,j))}(1-p_{\text{in}})^{(n_{ij}-a_{ij})B(i,j)}(1-p_{\text{out}})^{(n_{ij}-a_{ij})(1-B(i,j))}, \notag
\end{eqnarray}
where $B(i,j)=1$ if $i$ and $j$ are in the same block of $B$ and $B(i,j)=0$ otherwise.
We also note that the choice to view all within cluster edges (via $p_{\text{in}}$) and all between cluster edges (via $p_{\text{out}}$) interchangeably is a logical choice based on our prior understanding of the U.S.\ Senate and Supreme Court.
We discuss this further in Sections \ref{section:Senate-application} and \ref{section:supreme court-application}.

\subsection{Temporal network clustering}\label{section:temporal}

The Supreme Court dataset spans fifteen judicial terms (1990--2004), each giving forth an interaction array $A^t$ and a clustering $B^t$, $t=1990,\ldots,2004$.
We wish to model temporal changes to the clusterings $(B^t)_{t=1990,\ldots,2004}$ but in a way that is smooth with respect to short-term irregularities.
For this, we model each $A^t$, given $(B^s)_{s=1990,\ldots,2004}$, according to the Binomial stochastic blockmodel of Section \ref{section:types} but with success probability parameters $p^t=(p^t_{ij})_{1\leq i,j\leq n}$ varying with $t$.
To incorporate dependence over time, we now model $(B^t)_{t=1990,\ldots,2004}$ as a Markov chain on the space of partitions of $[n]$, where $n=9$ is the number of justices.

Generally, partitions with a small number of clusters relative to the sample size are most informative, and until recently there were no known partition-valued Markov chains with suitable properties for this application.
Using the {\em Ewens cut-and-paste chain} \citep{Crane2011a,Crane2014AOP}, we specify parameters $\alpha>0$, $k\geq2$ (the maximum number of clusters in each $B^t$) and we model a partition sequence $(B^t)_{t=0,1,\ldots}$ with transition probabilities
\begin{equation}\label{eq:tps}
P(B^{t+1}=\pi'\mid B^t=\pi;\alpha,k)=k^{\downarrow\#\pi}\prod_{b\in\pi}\frac{\prod_{b'\in\pi'}(\alpha/k)^{\uparrow\#(b\cap b')}}{\alpha^{\uparrow\#b}},
\end{equation}
where $\pi,\pi'$ are partitions of $[n]$, $\#\pi$ is the number of non-empty clusters of $\pi$, $\#b$ is the cardinality of cluster $b\in\pi$, $k^{\downarrow j}=k(k-1)\cdots(k-j+1)$, and $\alpha^{\uparrow j}=\alpha(\alpha+1)\cdots(\alpha+j-1)$.
This family of transition probabilities is reversible with respect to the Ewens--Pitman distribution with parameter $(-\alpha,k\alpha)$:
\begin{equation}\label{eq:EP}
P(B^0=\pi;\alpha,k)=\frac{k^{\downarrow\#\pi}}{(\alpha k)^{\uparrow n}}\prod_{b\in\pi}\alpha^{\uparrow\#b}.
\end{equation}

This class of Markov chains  has many properties that are suitable for the intended hidden Markov model application.
Most important for our applications below, any Markov chain $(B^t)_{t=0,1,\ldots}$ with initial distribution \eqref{eq:EP} and transition probabilities \eqref{eq:tps} is {\em exchangeable}, i.e.,  the sample can be relabeled arbitrarily without affecting the distribution of the sequence.
Since both the Poisson and Binomial stochastic blockmodels are {\em label equivariant}, i.e., the distribution of the data array $(A_{ij})_{1\leq i,j\leq n}$ under relabeling is unchanged provided the clustering parameter $B$ is relabeled in kind, their combination with the hidden Markov chain $(B^t)$ is unaffected by the arbitrary assignment of labels to individuals.
In the supreme court dataset below, the Court's membership changes during the period 1990--2004, meaning the sequence $(B^t)_{t=1990,\ldots,2004}$ does not represent partitions of the same set of individuals over time.
The above model is well equipped to handle this with an important {\em sampling consistency} property: given a Markov chain $(B^t)_{t=0,1,\ldots}$ on partitions of $[n]$ from the Ewens cut-and-paste chain, the restricted sequence $(B^t_{[m]})_{t=0,1,\ldots}$ obtained by removing individuals $m+1,\ldots,n$ from the sample is once again a Ewens cut-and-paste chain on partitions of $[m]$.
This sampling consistency property, therefore, allows us to model the temporal sequence $(A^t)_{t=1990,\ldots,2004}$ without any concerns.
Prior analyses of the Court, notably \citet{Sirovich2003PNAS} and \citet{ThurstoneDegan1951}, are restricted to short periods of time during which the Court's membership remained constant.

Reversibility is also a natural property since, although the arrow of time invariably moves toward the future, there is no logical mandate against analyzing the data in the reverse direction.
Moreover, since we seek to detect regime change, it is important that the model does not bias the sequence in any way.
Without knowledge to the contrary, we assume that each $B^t$ obeys the same marginal distribution, i.e., the chain evolves in equilibrium.
In this way, detected changes in $(B^t)_{t=1990,\ldots,2004}$ reflect meaningful information in the data instead of arbitrary defects in the model.

\section{Cluster analysis}\label{section:analysis}

All our analyses below proceed by optimizing an objective function, i.e., likelihood, posterior, or modularity measure, over the space of partitions of $[n]$.
Where scalar parameters, generically denoted $\theta$, are present, we can often compute unbiased, or asymptotically unbiased, estimates in closed form, which we profile out when searching for the optimal clustering.
Given an observed interaction array $A=a$, 
we write $g(B;\theta,a)$ as the generic objective function and we seek to solve 
\begin{equation}
\label{postmax}
\arg \max_{B} g(B;\hat{\theta}_B,a)
\end{equation}
where $\hat{\theta}_B$ is the maximum likelihood estimate of $\theta$ given $(B,a)$.

Similarly for temporal clustering, we seek the sequence $(B^t)_{t=0,1,\ldots,T}$ with the largest posterior probability.
This latter activity is, in general, quite computationally challenging; however, we leverage properties of the chosen model to mitigate these issues.
Using the stationarity of the hidden Markov chain for $(B^t)_{t=0,1,\ldots,T}$, we build up our estimated temporal clustering sequence sequentially.
We begin with the initial state $B^0$, which we equip with prior as in \eqref{eq:EP} with $k=2$ and $\alpha$ set to $1$---sensitivity analysis shows that our estimates are robust to this choice of $\alpha$---and we take the posterior mode based on the observed interactions in $A^0$ as our estimate  $\hat{B}^0$.
To estimate $B^{t+1}$, given $A^{t+1}$ and $(\hat{B}^{t},\ldots,\hat{B}^0)$, we use the conditional distribution in \eqref{eq:tps} from state $\hat{B}^t$ as our prior and again take our estimate $\hat{B}^{t+1}$ as the posterior mode.
Therefore, our estimate for $(B^t)_{t=0,1,\ldots,T}$ amounts to a slightly modified version of the search algorithm below.
A search over possible perturbations of the inferred sequence $(\hat{B}^t)_{t=0,1,\ldots}$ obtained in this way does not find any improvements.

\subsection{Randomized search algorithms}\label{section:algorithms}

When the sample size is moderate to large, the space of partitions is too big to search exhaustively during cluster detection.
To optimize the objective function \eqref{postmax}, we use the following randomized search algorithm which has been proven to efficiently search the space of partitions with a bounded number of blocks \citep{CraneLalley2012a} and has been used effectively in previous clustering applications \citep{Crane2014JASA,Crane2014Biometrika}.  For Newman-Girvens modularity, we employ the label-switching algorithm from \cite{BickelChen2009PNAS}.
The benefit of our algorithm over previous randomized algorithms, e.g., the split-and-merge algorithm used by \citet{BoothCasellaHobert2008}, is that we can restrict ours to only search over partitions with a maximum number of clusters, making the search much more efficient.

Our search algorithms iterate between local- and global-move Markov chains on the space of partitions.  
In all our applications in this paper, we fix the maximum number of clusters $k$.
(Note that in the case of the hidden Markov model above, $k$ here is the same value as in \eqref{eq:tps}.  They both correspond to the maximum number of clusters in $B$.)
To ensure the global moves do not suggest partitions with more than $k$ clusters, our algorithm proposes moves according to the transition probabilities in \eqref{eq:tps} with parameter $\tilde{\alpha}>0$ that is logically unrelated to the parameter $\alpha>0$ in the model.
Importantly, $k$ is only an upper bound on the number of clusters, so our choice of $k$ does not mandate exactly $k$ clusters as, e.g., $k$-means \citep{Lloyd1982} and Gaussian mixture models \citep{BanfieldRaftery1993}.
For $\tilde{\alpha}>0$, we recall the $\PE(-\tilde{\alpha},k\tilde{\alpha})$ distribution on partitions of $[n]$ from \eqref{eq:EP}.

\subsection{Global search: cut-and-paste algorithm}  
For $\tilde{\alpha}>0$ and $k=1,2,\ldots$, the Ewens cut-and-paste chain with parameter $(\tilde{\alpha},{k})$ evolves on partitions of $[n]$ with at most ${k}$ blocks according to the transition probabilities in \eqref{eq:tps}.
Here we describe how to efficiently generate transitions in this chain according to the {\em cut-and-paste} procedure.
  Let $\pi=\{b_1,\ldots,b_r\}$, $r=1,\ldots,k$, be the current state of the chain.
The next state is obtained as follows:
\begin{itemize}
	\item[(a)] Independently, each block $b_i$ is partitioned into $\tilde{\pi}^i$ according to \eqref{eq:EP} with parameter $(-\tilde{\alpha}/{k},\tilde{\alpha})$;
	\item[(b)] for each $i=1,\ldots,r$, the blocks of $\tilde{\pi}^i$ are labeled uniformly without replacement in $\{1,\ldots,{k}\}$; and
	\item[(c)] the next state $\pi'$ is obtained by aggregating blocks in (b) with the same label and then removing the labels.
\end{itemize}

The most attractive feature of the cut-and-paste chain
is that it assigns strictly positive probability to any transition in the search space (and therefore moves around the space quickly).
This intuition is supported by rigorous proof that it converges to its stationary distribution in $O(\log n)$ steps
\citep{CraneLalley2012a}, where $n$ is the number of vertices.

\subsection{Local search: cocktail algorithm} 
For $\tilde{\alpha}>0$ and $k=1,2,\ldots$, the cocktail algorithm evolves on partitions by updating
one element at a time.  Let $\pi$ be the current state of the
chain.  First, an element $u\in[n]$ is sampled uniformly at random and removed from 
$\pi$ to obtain $\pi_{[n]\setminus u}$.  Given $\pi_{[n]\setminus u}$, the removed
element $u$ is reinserted into $\pi_{[n]\setminus u}$ according to the seating rule 
of the $(-\tilde{\alpha},k\tilde{\alpha})$-Chinese restaurant process:
\[\pr(u\mapsto b\mid\pi_{[n]\setminus u})\propto\left\{\begin{array}{cc} \#b-\tilde{\alpha},& b\in\pi_{[n]\setminus u}\\
k\tilde{\alpha}+\tilde{\alpha}\#\pi_{[n]\setminus u},& b=\emptyset.
\end{array}\right.\]
At each step, this chain is distributed according to \eqref{eq:EP} and, therefore, is confined to partitions with at most $k$ blocks.

By iterating between the local and global chains, our search algorithm explores
the partition space for local and global maxima.  To effectively use this algorithm, we 
take a step in the global chain followed by a prespecified number of moves in the cocktail
algorithm.  We accept all moves in the global chain, and we accept moves in the cocktail chain according to the Metropolis--Hastings
algorithm.  This choice reflects our observation that
local maxima often occur only a few steps away from partitions with low likelihood, and so 
rejecting global moves can be counter-productive to search.  
The efficiency of this method is apparent in our application for the senate dataset (Section \ref{section:Senate-application}), where we use it to search over all partitions of 100 senators into at most two blocks, starting in a randomly chosen starting state.
This space consists of $2^{99}\approx6.3\times10^{29}$ partitions, but our algorithm converges quickly to the right answer on a laptop computer.

\section{Applications}\label{section:applications}

\subsection{Karate club}\label{section:karate-application}
\begin{figure}
\centering
\mbox{\subfigure{\includegraphics[width=6cm]{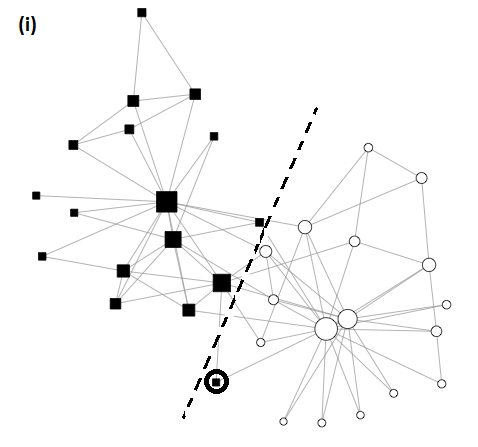}}\quad
\subfigure{\includegraphics[width=6cm]{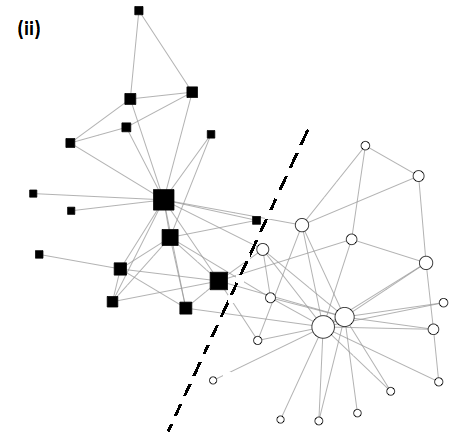} }}
		\includegraphics[width=8cm]{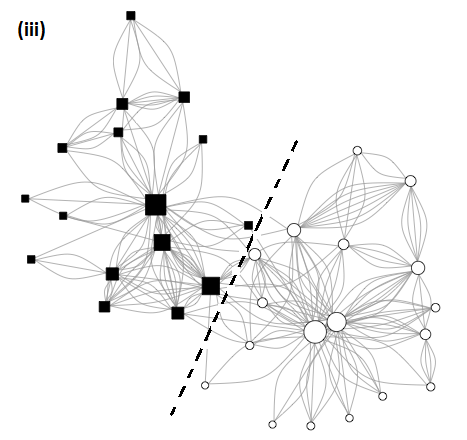}
	\caption{Inferred clusterings for the karate club dataset.  Dotted line in each panel marks the separation of clusters according to the analysis in \citet{Zachary1977}.  Black squares and white circles indicate the two different clusters inferred by the chosen method in each panel.  (i) Inferred community structure using Newman--Girvan modularity on projected network with cutoff $c=0$.  (ii) Inferred community structure using Newman--Girvan modularity on projected network with cutoff $c=1$.  (iii) Inferred community structure using Poisson stochastic blockmodel on full interaction array.  The circled individual in Panel (i) is inconsistently classified by Newman--Girvan in (i) and (ii). }
	\label{figure:1}
\end{figure}

We fit the Poisson stochastic blockmodel to the full interaction array from the karate club dataset.
To best compare with  previous methods, we found the best fit with at most two clusters and two parameters $\lambda_{\text{in}},\lambda_{\text{out}}>0$ for within- and between-cluster intensities.
Our inferred clustering in Figure \ref{figure:1}(iii) (based on maximum likelihood for \eqref{eq:Poisson-SBM}) is correct according to the analysis in \citet{Zachary1977}.
Under our model the likelihood for this correct clustering is $-348.26$ with estimated intensities $(\hat{\lambda}_{\text{in}},\hat{\lambda}_{\text{out}})=(0.615,0.066)$, versus $-349.81$ with $(\hat{\lambda}_{\text{in}},\hat{\lambda}_{\text{out}})=(0.618,0.066)$ for the clustering found by the degree-corrected stochastic blockmodel and Newman--Girvan modularity in Figure \ref{figure:1}(i).
We point out that our analysis does not contradict the findings of \citet{KarrerNewman2011}, who report that the Poisson stochastic blockmodel without degree correction ``fails to split the network into the known factions.''
That conclusion is based on fitting the generic Poisson stochastic blockmodel with different within-cluster intensities $\lambda_{\text{in},1},\lambda_{\text{in},2}>0$ to the {\em projected} network data.
Given this flexibility, it is not surprising that the clustering divides the group into high- and low-degree individuals, as that inference also has a reasonable and clear interpretation in terms of the given model.

In our view, it is not fair to conclude that the Poisson stochastic blockmodel {\em failed} in this instance, as Karrer and Newman claim.
By separating the highly connected individuals into a single cluster, the detected clustering does accurately extract a low resolution overview of the network.
That this does not coincide with the desired ``true'' clustering suggests only that the specified model was not set up to detect two clusters of similar size and characteristics.
The network reported by Zachary has the feature that the two clusters exhibit similar characteristics, and our choice of $\lambda_{\text{in},1}=\lambda_{\text{in},2}$ reflects our interest in detecting the best such clustering.
This constraint is also consistent with our prior understanding of the karate club dataset, for which we have no {\em a priori} reason to expect within-cluster social interactions of different clusters to be substantially different.
Our constraint, therefore, allows us to pool information from the two clusters, obtaining the correct split.

As \citet{KarrerNewman2011} report, the degree-corrected model returns a better clustering than the Poisson stochastic blockmodel with three parameters $\lambda_{\text{in},1},\lambda_{\text{in},2},\lambda_{\text{out}}>0$ but only after introducing a new degree-correction parameter for each of the thirty-four individuals in the network.
Instead of decreasing the number of parameters to two and recovering the correct clustering, the degree-corrected model increases the number of parameters to thirty-six and still incorrectly classifies one individual.

\subsubsection{Fitting the projected network}
While our analysis of the full interaction data (the multigraph in Figure \ref{figure:1}(iii)) recovers the correct clustering as reported by \citet{Zachary1977},  the other methods misspecify one individual.
Even more curious is the behavior of the Newman--Girvan method under different choices of projection.
In the {\em standard} karate club projection, that obtained by cutoff $c=0$,  Newman--Girvan incorrectly specifies one individual; however, if the projection uses cutoff $c=1$, then Newman--Girvan finds the correct clustering.
These results are shown in Figure \ref{figure:1}(i) and (ii), with the misspecified individual circled in panel (i).

A closer look at the data explains the discrepancy.
The misclassified individual is connected to the most highly connected vertices in both clusters, i.e., one interaction with the highest degree vertex in the left cluster (black squares) and two interaction with the highest degree vertex in the right cluster (white circles).
The standard projection with $c=0$ records both of these as a single edge between these individuals in both clusters, leading the Newman--Girvan algorithm astray.
When projecting with $c=1$, however, the single interaction with the black cluster is thresholded out, leaving only one interaction to the high degree vertex in the white cluster.
This highlights, on a small scale, that the arbitrary choice of projection does have an effect on the inferred clustering and should raise concerns about inferences that do not account, or cannot account, for the projection operation.
Comparison of the weighted interaction network in Figure \ref{figure:1}(iii) with the projected networks in Figure \ref{figure:1}(i) and (ii) demonstrates visually the amount of information discarded in projecting to an unweighted network.
The temperamental  nature of inferences based on projected networks is much more pronounced on the senate dataset, as we now show.



\subsection{Senate voting}\label{section:Senate-application}

\begin{figure}
\centering
\mbox{\subfigure{\includegraphics[width=7cm]{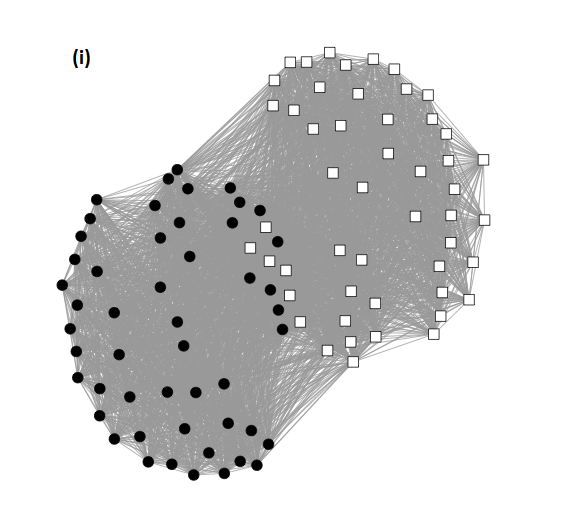}}\quad
\subfigure{\includegraphics[width=7cm]{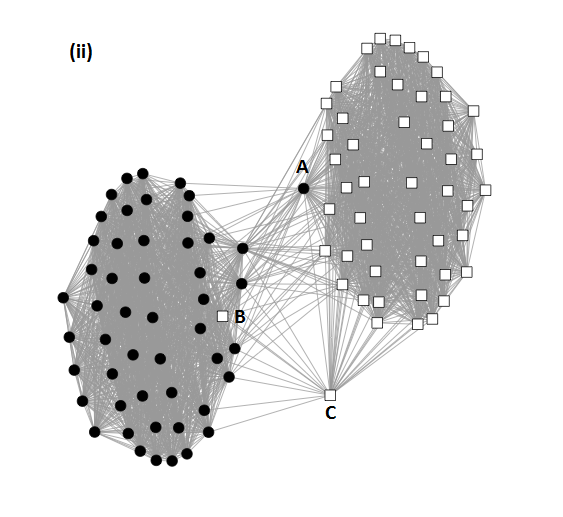} }}
	\includegraphics[width=7cm]{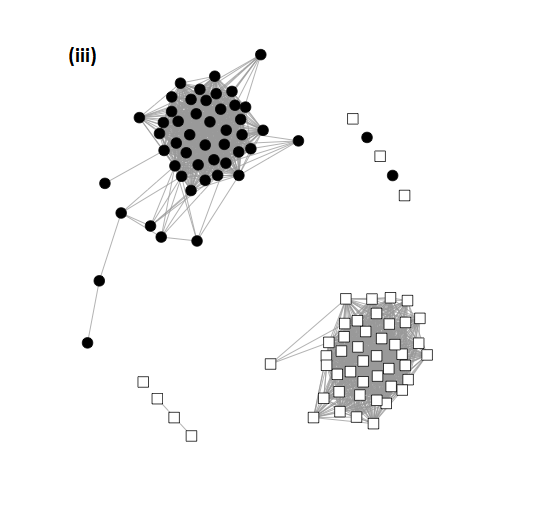}	\label{figure:2}
	\caption{Projected networks obtained from senate voting dataset for cutoff values chosen as the (i) 30th, (ii) 50th, and (iii) 70th percentile of ratios $V_{ij}/N_{ij}$ in the interaction array.  (See Section \ref{section:senate} for further explanation.) 
In all panels, Republicans are indicated by white squares and Democrats by black dots.
In panel (ii), the labeled vertices are (A) Zell Miller (D-GA), (B) Jim Jeffords (R-VT), and (C) Lincoln Chafee (R-RI).}
\end{figure}

\begin{table}
\caption{\label{table:percentile} Performance of Newman--Girvan modularity for different cutoff values in the projected senate dataset.
Misclassified individuals are those assigned to the wrong cluster according to NG modularity.
Nonclassified individuals are those who cannot be assigned to a cluster because the projection causes isolated vertices.}
\fbox{%
\begin{tabular}{l|ccccccccccc}
percentile cutoff&  20 & 25 & 30 & 35 & 40 & 45 & 50 & 55 & 60 & 65 & 70 \\\hline
misclassified & 12  & 8 & 4 & 4  & 4 & 0  & 0 & 0 & 1 & 6 & 2\\
nonclassified & 0 & 0 & 0 & 0 & 0 & 0 & 0 & 0 & 2 & 4 & 7\\\hline
total & 12 & 8 & 4 & 4 & 4 & 0 & 0 & 0 & 3 & 10 & 9
\end{tabular}}
\end{table}

While studying clustering methods from categorical data, \cite{Crane2014JASA} analyzed voting data from the 107th U.S.\ Senate.
The U.S.\ Senate consists of 100 elected individuals, each of whom vote {\em yea} or {\em nay} on a series of amendments.
Using a three-parameter extension of the Ewens--Pitman two-parameter partition model \citep{Ewens1972,PermanPitmanYor}, \cite{Crane2014JASA} detected a partition into two equally sized clusters, but with one Democrat and one Republican defecting into the opposing cluster.
Here the interaction array $A$ contains information about votes of different types, as we discussed in Section \ref{section:types}.
We fit the Binomial stochastic blockmodel with parameters $p_{\text{in}},p_{\text{out}}\in[0,1]$, as we have no expectation of different qualitative behavior between clusters.
Using maximum likelihood estimation in combination with the randomized search algorithms from Section \ref{section:analysis}, our analysis from the Binomial stochastic blockmodel correctly finds the clustering from \citet{Crane2014JASA} (log-likelihood: $-57271.15$ with $\hat{p}_{\text{in}}= 0.858$ and $\hat{p}_{\text{out}}=0.529$).
(For comparison to the most obvious candidate, the clustering of senators along party lines returns a likelihood of $-68019.87$ with $\hat{p}_{\text{in}}=0.853$ and $\hat{p}_{\text{out}}=0.533$.)

Upon further inspection, the inferred clustering detects a reasonable departure from the expected clustering along party lines: the Democrat in question, Zell Miller (D-GA), was a vocal supporter of Republican president George W.\ Bush and was regarded by many as a traitor to his party; the Republican in question, Jim Jeffords (R-VT), switched to the Democratic party later in the term.


\subsubsection{Fitting the projected network}
Figure \ref{figure:2} shows the Senate network under various choices of projection based on the ratios $V_{ij}/N_{ij}$ for each pair.
This ratio gives the overall frequency of agreement between senators $i$ and $j$ for bills on which they both voted.
Panel (i) shows that the projected network with cutoff value at the 30th percentile of ratios ruins much of the structure in the data, while panel (iii) shows that the projected network with cutoff value at the 70th percentile of ratios leaves certain vertices isolated and, therefore, unable to be classified.
Panel (ii) shows that the cutoff chosen as the 50th percentile separates the two clusters pretty well, with Miller (labeled vertex (A)) and Jeffords (labeled vertex (B)) aligned in the cluster of the opposite party.
Vertex (C) is Lincoln Chafee (R-RI) who has strong ties to both parties and, in fact, has since joined the Democratic party.

Table \ref{table:percentile} details the performance of the Newman--Girvan modularity for different choices of cutoff.  
The misclassifications are due to a flat modularity across several alternatives to the ``true'' clustering~$B^\star$; for the 30th
percentile, for example, there is a clustering with 4 misclassified nodes but identical NG-modularity to $B^\star$.  
The Newman--Girvan modularity is able to correctly identify the clusters with cutoffs between the 45th and 55th percentile, but Figure \ref{figure:2}(ii) illustrates that this is due entirely to the choice of projection.
Upon projecting to the network in panel (ii), we could determine the clusters by visual inspection, without any need to run an algorithm.

While not explicitly discussed by \cite{BickelChen2009PNAS}, Newman--Girvan modularity can be used on the weighted
matrix~$T$ such that $T_{ij} = t(A_{ij}) = V_{ij}/N_{ij}$.  
In the case of the senate, the true clustering does maximize NG modularity based on the weighted matrix $T$, but there are several other local optima with the same modularity.
In this case, NG modularity cannot confirm nor deny the true clustering.



%
%

\subsection{Temporal clustering in the Supreme Court}\label{section:supreme court-application}

\begin{figure}
\centering
	\includegraphics[width=12cm]{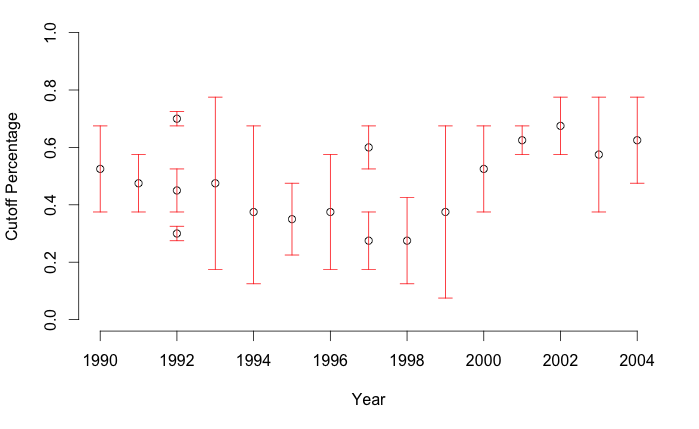}	\label{figure:3}
	\caption{Plot of cutoff ranges for which the Newman--Girvan modularity recovers the correct Supreme Court clustering.  The circle within each range represents the median cutoff value.}

\end{figure}

For inferring temporally-varying community structure, we consider the collection of interaction arrays $(A^t)_{t=1990,\ldots,2004}$ from the U.S.\ Supreme Court. 
A notable feature of this dataset is that the Court's membership is not constant during this time, with new additions of Clarence Thomas, Ruth Bader Ginsburg, and Stephen Breyer in 1991, 1993, and 1994, respectively.
Our chosen Ewens cut-and-paste chain as a hidden Markov model for the clustering sequence $(B^t)_{t=1990,\ldots,2004}$ easily handles this by its sampling consistency property.
The key feature of this dataset from the viewpoint of model verification focuses on the cluster membership of Justice David Souter, who was appointed to the Court in 1990 by Republican President George H.W.\ Bush.
Most legal scholars \citep{Irons,Toobin2008} point out a shift in his judicial philosophy, from initially conservative to more liberal in 1993.
Prior quantitative analyses of the Court detect this change in ideology, e.g., by the ideal points method of \citet{MartinQuinn2002}.
Those analyses generally incorporate much more information about Supreme Court jurisprudence, such as details about specific cases, but our inferred clusterings in Table \ref{table:sc-time} obtain the same inference with only the interaction array data $(A^t)_{t=1990,\ldots,2004}$.

As an algorithmic method, Newman--Girvan modularity is not properly equipped to handle dependence over time.
To illustrate the benefit of smoothing temporal irregularities with the hidden Markov chain, we fit the Newman--Girvan modularity independently to projection data from each term.
Although each term has a window of percentile cutoffs for which the algorithm detects the true clustering, these windows move substantially from term to term, and visual inspection of the figure says that no single choice of threshold yields the correct clustering for every term.

\begin{table}
\caption{\label{table:sc-time} Estimated ideological cluster sequence $(B_t)_{t=1990,\ldots,2004}$ from the Binomial stochastic blockmodel with temporally varying community structure modeled with the Ewens cut-and-paste chain as a hidden Markov model.
Black and white circles indicate cluster membership within each term, with missing classifications indicating that the justice was not on the Court for the given term.
Note that our method correctly detects the ideological shift of David Souter between the 1992 and 1993 terms.}
\centering
\fbox{%
\begin{tabular}{l|ccccccccccccccccc}
Justice &  1990 &91 & 92 & 93 & 94 & 95 & 96 & 97 & 98 & 99 & 00 & 01 & 02 & 03 & 04 & \\\hline
White& $\circ$ &  $\circ$ & $\circ$ &&&&&&&&&&&&\\
Marshall & $\bullet$ & &&&&&&&&&&&&\\
Blackmun & $\bullet$& $\bullet$ & $\bullet$ & $\bullet$ &&&&&&&&&&&\\
Rehnquist & $\circ$&$\circ$ & $\circ$ & $\circ$ & $\circ$ & $\circ$& $\circ$& $\circ$& $\circ$& $\circ$& $\circ$& $\circ$& $\circ$& $\circ$& $\circ$\\
Stevens & $\bullet$&$\bullet$ & $\bullet$ & $\bullet$ & $\bullet$ & $\bullet$ & $\bullet$ & $\bullet$ & $\bullet$ & $\bullet$ & $\bullet$ & $\bullet$ & $\bullet$ & $\bullet$ & $\bullet$ \\
O'Connor &$\circ$&$\circ$ & $\circ$ & $\circ$ & $\circ$ & $\circ$& $\circ$& $\circ$& $\circ$& $\circ$& $\circ$& $\circ$& $\circ$& $\circ$& $\circ$\\
Scalia & $\circ$&$\circ$ & $\circ$ & $\circ$ & $\circ$ & $\circ$& $\circ$& $\circ$& $\circ$& $\circ$& $\circ$& $\circ$& $\circ$& $\circ$& $\circ$\\
Kennedy & $\circ$  &$\circ$ & $\circ$ & $\circ$ & $\circ$ & $\circ$& $\circ$& $\circ$& $\circ$& $\circ$& $\circ$& $\circ$& $\circ$& $\circ$& $\circ$\\
Souter & $\circ$&$\circ$ & $\circ$ & $\bullet$ & $\bullet$ &   $\bullet$ & $\bullet$ & $\bullet$ & $\bullet$ & $\bullet$ & $\bullet$ & $\bullet$ & $\bullet$ & $\bullet$ & $\bullet$ \\ 
Thomas &&$\circ$ & $\circ$ & $\circ$ & $\circ$ & $\circ$& $\circ$& $\circ$& $\circ$& $\circ$& $\circ$& $\circ$& $\circ$& $\circ$& $\circ$ \\
Ginsburg& & & & $\bullet$ & $\bullet$ & $\bullet$ & $\bullet$ & $\bullet$ & $\bullet$ & $\bullet$ & $\bullet$ & $\bullet$ & $\bullet$ & $\bullet$ & $\bullet$ \\
Breyer && &&  & $\bullet$ & $\bullet$ & $\bullet$ & $\bullet$ & $\bullet$ & $\bullet$ & $\bullet$ & $\bullet$ & $\bullet$ & $\bullet$ & $\bullet$ \\
\\\hline\hline\\
\end{tabular}}
\end{table}
\normalsize

\section{Concluding remarks}

If considered within the proper context, classical methods may have potential in network applications, as they can be readily
built upon for more advanced statistical inference.  
Our analysis of temporal variation ideological clusters within the U.S.\ Supreme Court embodies a healthy cross-fertilization with ideas in the applied and theoretical probability literature.
Combining this with the straightforward Binomial stochastic blockmodel of Section \ref{section:types}, we correctly detect David Souter's ideological shift after his third term.
By comparison, the Newman--Girvan modularity, or any algorithmic method we know of, cannot adequately deal with temporal variation in the underlying clustering.
In picturesque fashion, Figure \ref{figure:3} points out that no single choice of cutoff yields the true latent clusters for all years in the supreme court dataset.

Proceeding from first principles, we have found that straightforward modifications of standard statistical methods perform just as well as prevailing clustering algorithms and blockmodel approaches for detecting communities in interaction datasets.
At some level, our analysis reiterates the common sense notion that throwing away data adversely affects inference.
At a deeper level, it is a call to think deeply about the role sampling plays in any network inference, an issue that has been raised by some \citep{CraneDempsey2015,LeeKimJeong2006,WillingerAldersonDoyle2009} but is largely ignored in the methodology literature.
If nothing else, our investigation calls special attention to the precarious behavior of the Newman--Girvan modularity with respect to the mechanism by which network data is sampled from interaction counts.

\section*{Acknowledgement}
H. Crane is partially supported by NSF grants DMS-1308899 and CNS-1523785 and NSA grant H98230-13-1-0299.

\bibliography{network-refs}
\bibliographystyle{chicago}

\end{document}